\documentclass[a4paper,11pt]{article}
\pdfoutput=1 % if your are submitting a pdflatex (i.e. if you have
\usepackage{jcappub} % for details on the use of the package, please
\usepackage[T1]{fontenc} % if needed
\usepackage{bm}
\usepackage{indentfirst}
\usepackage{amsmath}
\usepackage{graphicx}
\usepackage{amssymb}
\usepackage{subfigure}
\usepackage{psfrag}
\usepackage{tensor}
\usepackage{braket}
\usepackage{multirow}
\usepackage{verbatim}
\usepackage[normalem]{ulem}
\usepackage{cancel}
\usepackage{nameref}
\usepackage{hyperref}
\usepackage{cleveref}
\usepackage{orcidlink}
\title{\boldmath Scalar induced gravitational waves in PV symmetric teleparallel gravity with a non-minimally coupled boundary term}

\author[a,1]{Fengge Zhang \orcidlink{0000-0001-7365-5057}\note{Corresponding author}}
\affiliation[a]{Institute for Gravitational Wave Astronomy, Henan Academy of Sciences, Zhengzhou, Henan 450046, China}
\emailAdd{zhangfengge@hnas.ac.cn}
\abstract{The question of whether parity violation occurs in gravitational interactions has recently attracted considerable attention. In this paper, we investigate the scalar induced gravitational waves (SIGWs) in symmetric teleparallel gravity incorporating a parity-violating (PV) term and a non-minimally coupled boundary term. The presence of this non-minimally coupled boundary term helps to avoid the strong coupling problem inherent in PV symmetric teleparallel gravity, thereby enabling our study of SIGWs in this PV symmetric teleparallel gravity. We numerically calculate the SIGWs generated during the radiation-dominated era and compute the corresponding energy density spectrum with a monochromatic primordial power spectrum. The resulting energy density spectrum of SIGWs exhibits significant deviations from the predictions of general relativity, particularly at high frequencies. This distinctive feature is detectable by future space-based gravitational wave detectors like LISA, TianQin, and Taiji.}

\begin{document}
\maketitle
\flushbottom

\section{Introduction}\label{introduction}
As the standard theory of gravity, general relativity (GR) has achieved profound success in describing astronomical and cosmological phenomena. Nevertheless, GR faces both theoretical and observational challenges--including spacetime singularities and the early-time and late-time accelerated expansion of the universe \cite{Starobinsky:1980te,Guth:1980zm,SupernovaSearchTeam:1998fmf,SupernovaCosmologyProject:1998vns}, which motivate the exploration of modified gravity theories. Testing gravity beyond GR remains one of the central pursuits in physics. 

The direct detection of gravitational waves (GWs) generated from compact binary mergers by LIGO-Virgo collaborations \cite{Abbott:2016blz,TheLIGOScientific:2017qsa,Abbott:2017oio,Abbott:2017vtc,LIGOScientific:2018mvr,Abbott:2020uma} has opened a new avenue for probing the nature of gravity in the strong gravitational field and nonlinear regimes, while also marking the dawn of multi-messenger astronomy. In addition to these astrophysical sources, primordial GWs, as a component of stochastic gravitational wave background (SGWB), carry valuable information about the early universe. However, primordial GWs have not yet been detected in cosmic microwave background (CMB) observations, with the tensor-to-scalar ratio constrained to $r<0.036$ at $95\%$ confidence \cite{BICEP:2021xfz,Planck:2018jri}. Beyond linear order, scalar induced gravitational waves (SIGWs),  generated through nonlinear interactions between tensor and scalar perturbations {\cite{Tomita:1967wkp,Matarrese:1992rp,Matarrese:1993zf,Matarrese:1997ay,Carbone:2004iv,Baumann:2007zm,Ananda:2006af,Saito:2008jc,Mangilli:2008bw,Martineau:2007dj,Bartolo:2007vp,Boubekeur:2008kn,Assadullahi:2009jc,Alabidi:2012ex,Kawasaki:2013xsa,Domenech:2019quo,Hwang:2017oxa,Kohri:2018awv,Espinosa:2018eve,Lu:2020diy,Domenech:2021ztg}, can be large enough to be detected by future space-based GW detectors such as Laser Interferometer Space Antenna (LISA) \cite{Danzmann:1997hm,LISA:2017pwj,Robson:2018ifk,LISACosmologyWorkingGroup:2025vdz,LISACosmologyWorkingGroup:2024hsc,Escriva:2024ivo,Perna:2024ehx,Kugarajh:2025rbt}, TianQin \cite{Luo:2015ght,Gong:2021gvw} and Taiji \cite{Hu:2017mde,Ruan:2018tsw}, as long as the primordial scalar perturbations are significantly enhanced on small scales \cite{Domenech:2019quo,Sato-Polito:2019hws,Lu:2019sti,Fu:2019vqc,Lin:2020goi,Zhang:2021vak,
Zhang:2021rqs,Yi:2022ymw,Lin:2021vwc}. Moreover, the SIGWs are also a potential signal source for pulsar timing arrays \cite{NANOGrav:2023hvm,EPTA:2023xxk,Yi:2023mbm,Liu:2023ymk,Chen:2024fir}.

Parity-violating (PV) gravity has recently attracted considerable attention \cite{Jackiw:2003pm,Horava:2009uw,Crisostomi:2017ugk,Gao:2019liu,Hu:2021bbo,Hu:2024hzo}. Parity symmetry implies that the laws of physics are invariant under a left-right directional flip.
Established in the weak interaction \cite{Lee:1956qn,Wu:1957my}, parity violation may also leave imprints on gravity. Recent cosmological observations, including measurements of the galaxy trispectrum and cross-correlation of $E$ and $B$ modes polarization, have provided tentative hints of parity violation in our universe \cite{Philcox:2022hkh,Hou:2022wfj,Minami:2020odp,Eskilt:2022cff}. In PV gravity, the left and right-hand polarization modes of GWs behave differently due to the PV effects, which results in distinctive features such as the velocity and amplitude birefringence, as well as circular polarization \cite{Takahashi:2009wc,Wang:2012fi,Zhu:2013fja,Cannone:2015rra,Zhao:2019szi,Qiao:2019hkz,Qiao:2021fwi,Gong:2021jgg,Wang:2012fi,Zhang:2025kcw,Guo:2025bxz,Feng:2024yic}. While individual planar interferometers, such as LISA and Taiji, respond only to the intensity Stokes parameter $I = \mathcal{P}^{R}_{h} + \mathcal{P}^{L}_{h}$ for an isotropic SGWB, the LISA--Taiji network can also respond to the circular polarization parameter $V = \mathcal{P}^{R}_{h} - \mathcal{P}^{L}_{h}$, which characterizes the amplitude asymmetry between right- and left-handed GW power spectra \cite{Seto:2020zxw,Orlando:2020oko,Caporali:2025dyf}.
Consequently, parity violation in the tensor sector can, in principle, be decoded from the two-point correlation function (power spectrum) of GWs. In contrast, for scalar perturbations, the lowest-order correlation function that can extract PV information is the four-point function (trispectrum) \cite{Liu:2019fag,Niu:2022fki,Cabass:2022rhr,Creque-Sarbinowski:2023wmb,Garcia-Saenz:2023zue}.

Within Riemannian geometry, Chern-Simons (CS) gravity represents the simplest PV extension \cite{Jackiw:2003pm} and has been extensively studied in cosmology, linear GWs and SIGWs \cite{Lue:1998mq,Alexander:2009tp,Gluscevic:2010vv,Myung:2014jha,Nishizawa:2018srh,Bartolo:2018elp,Zhang:2022xmm,Feng:2023veu,Nojiri:2025jhh}. However, CS gravity suffers from theoretical pathologies. The higher-derivative equations of motion (EOM) lead to Ostrogradsky instabilities. Moreover, one of the gravitational wave polarization modes becomes a ghost when its momentum exceeds the characteristic PV energy scale \cite{Dyda:2012rj}, thus CS gravity can be viewed only as a low-energy effective field theory. In addition, CS gravity was also generalized to ghost-free chiral scalar-tensor gravity by including the higher derivatives of the coupling scalar field \cite{Crisostomi:2017ugk}. Using the method of spatially covariant gravity, the authors proposed a PV scalar-tensor theory with up to four derivatives that is free of Ostrogradsky and ghost instabilities \cite{Hu:2024hzo}.

Teleparallel geometry provides an alternative framework for constructing PV gravity theories \cite{Nieh:1981ww,Cai:2015emx,Cai:2021uup,Langvik:2020nrs,Li:2020xjt,Li:2021mdp,Conroy:2019ibo,Pagani:2015ema,Chen:2022wtz,Gialamas:2022xtt,Bahamonde:2021gfp,DAmbrosio:2021pnd,Papanikolaou:2022hkg,Tzerefos:2023mpe,Heisenberg:2023lru,Zhang:2023scq,Yu:2024drx,Zhang:2024vfw,Nojiri:2024zab}. In teleparallel gravity, spacetime is flat, and gravity is described by the non-metricity tensor $Q_{\rho\mu\nu}$ and/or torsion $T^{\rho}_{\ \mu\nu}$.

A simple and ghost-free PV model within the symmetric teleparallel framework was proposed by appending the PV term $\widetilde{Q}Q = \varepsilon_{\mu\nu\rho\sigma}Q^{\mu\nu}_{\phantom{\mu\nu}\lambda}Q^{\rho\sigma\lambda}$ to the Einstein–Hilbert equivalent Lagrangian \cite{Li:2021mdp}. The linear cosmological perturbations of this model have also been studied \cite{Li:2021mdp,Cai:2021uup}. However, when the nonlinear perturbations are taken into account, this simplest PV symmetric teleparallel gravity exhibits inconsistencies due to the strong coupling problem. This strong coupling problem arises because the connection perturbation modes participate in nonlinear interactions without independent dynamical EOMs, thereby precluding reliable SIGW calculations in this PV symmetric teleparallel gravity. 

In our previous work \cite{Zhang:2023scq}, we addressed this problem by replacing the symmetric teleparallel equivalent Einstein–Hilbert Lagrangian with a general linear combination of non-metricity tensor quadratic monomials. Although this approach successfully avoids the strong coupling problem, it introduces five free parameters. Even after imposing the no-higher-order-derivatives constraint, two independent parameters remain.

In this work, we implement an alternative approach to avoid the strong coupling problem. As noted previously, the strong coupling problem in model \cite{Li:2021mdp} originates from the absence of the linear EOMs for connection perturbations. To resolve this, we introduce a non-minimally coupled boundary term that naturally incorporates the connection perturbations into the quadratic action, providing them with linear dynamics and thereby eliminating the strong coupling problem. Then we derive and numerically solve the linear perturbation equations during radiation-dominated era and subsequently compute the energy density spectrum of SIGWs.

This paper is organized as follows. In section \ref{sec2}, we revisit the symmetric teleparallel gravity and introduce our model. In section \ref{sec3}, we present the EOMs for the background evolution and the linear scalar perturbations, subsequently solving them during the radiation-dominated era. In section \ref{sec4}, we compute both the power spectra of the SIGWs and the corresponding energy density spectrum of SIGWs with the monochromatic power spectrum of primordial curvature perturbation. Our conclusions are presented in section \ref{seccon}. The linear, quadratic and cubic actions are collected in appendix \ref{pactions}, and the detailed derivation of the SIGW calculation is provided in appendix \ref{calculationgws}.

\section{Revisiting the symmetric teleparallel gravity}\label{sec2}

In symmetric teleparallel gravity, the affine connection satisfies vanishing curvature and torsion:
\begin{equation}\label{curvature}
R^{\mu}_{\ \nu\rho\sigma}=\partial_{\rho}\Gamma^{\mu}_{\ \nu\sigma}-\partial_{\sigma}\Gamma^{\mu}_{\ \nu\rho}+\Gamma^{\mu}_{\ \alpha\rho}\Gamma^{\alpha}_{\ \nu\sigma}-\Gamma^{\mu}_{\ \alpha\sigma}\Gamma^{\alpha}_{\ \nu\rho}=0,
\end{equation}
\begin{equation}\label{torsion}
T^{\mu}_{\ \nu\rho}=\Gamma^{\mu}_{\ \rho\nu}-\Gamma^{\mu}_{\ \nu\rho}=0.
\end{equation}
The gravitational interaction is governed by the non-metricity tensor defined as
\begin{equation}
Q_{\rho\mu\nu}=\nabla_{\rho}g_{\mu\nu}=\partial_{\rho}g_{\mu\nu}-\Gamma^{\sigma}_{\ \rho\mu}g_{\sigma\nu}-\Gamma^{\sigma}_{\ \rho\nu}g_{\sigma\mu},
\end{equation}
where $g_{\mu\nu}$ denotes the spacetime metric and $\nabla$ represents the covariant derivative. 

The coefficients of an affine connection that maintains a flat \eqref{curvature} and torsionless \eqref{torsion} spacetime take the following general form \cite{BeltranJimenez:2017tkd,DAmbrosio:2020nqu,Heisenberg:2023lru}
\begin{equation}\label{cy}
\Gamma^{\rho}_{\ \mu\nu}=\frac{\partial x^{\rho}}{\partial \xi^{\sigma}}\partial_{\mu}\partial_{\nu}\xi^{\sigma},
\end{equation}
where $\xi^{\mu}(x)$ are four general scalar functions of coordinates $x^\mu$. Choosing $\xi^{\mu}(x)=x^{\mu}$ yields $\Gamma^{\rho}_{\ \mu\nu}=0$, corresponding to the coincident gauge. This gauge choice significantly simplifies the calculations and is widely adopted in symmetric teleparallel gravity studies.

The model proposed in \cite{Li:2021mdp} is given by the action 
\begin{equation}\label{action1}
S_g=\int \mathrm{d}^4x\sqrt{-g}\left(\frac{\mathbb{Q}}{2}-g_2(\varphi) \widetilde{Q}Q\right)+\int \mathrm{d}^4x\sqrt{-g}\left(\frac{1}{2}g^{\mu\nu}\partial_{\mu}\varphi\partial_{\nu}\varphi-V(\varphi)\right),
\end{equation}
where the non-metricity scalar
\begin{equation}\label{qrc}
	\mathbb{Q}=\frac{1}{4}Q_{\rho\mu\nu}Q^{\rho\mu\nu}-\frac{1}{2}Q_{\rho\mu\nu}Q^{\mu\nu\rho}-\frac{1}{4}Q^{\alpha}Q_{\alpha}+\frac{1}{2}\Bar{Q}^{\alpha}Q_{\alpha}=-\mathring{R}-B,
\end{equation}
corresponds to the teleparallel equivalent Einstein-Hilbert Lagrangian.
The boundary term $B$ is defined as,
\begin{equation}
B=\mathring{\nabla}_{\alpha}\left(Q^{\alpha}-\Bar{Q}^{\alpha}\right),
\end{equation}
and
\begin{equation}
Q_{\mu}=Q_{\mu\ \alpha}^{\ \alpha},\qquad \bar{Q}^{\mu}=Q_{\alpha}^{\ \mu\alpha},
\end{equation}
$\mathring{\nabla}$ is the metric-compatible covariant derivative and $\mathring{R}$ in Eq. \eqref{qrc} is the Ricci scalar computed from the Levi-Civita connection. 

The PV term takes the following form \cite{Conroy:2019ibo}
\begin{equation}
\widetilde{Q}Q=\varepsilon^{\mu\nu\rho\sigma}Q_{\mu\nu\alpha}Q_{\rho\sigma}^{\ \ \alpha},
\end{equation}
where $\varepsilon^{\mu\nu\rho\sigma}=\epsilon^{\mu\nu\rho\sigma}/\sqrt{-g}$ is the Levi-Civita tensor, with $\epsilon^{\mu\nu\rho\sigma}$ the antisymmetric symbol.
The scalar field $\varphi$ in action \eqref{action1} effectively describes the matter filled in the universe.

Although both the PV term and the boundary term involve the connection contributions, only the PV term appears in the system \eqref{action1}. Furthermore, in a Friedmann-Robertson-Walker (FRW) universe, the PV term does not contribute to the EOMs of the background and linear perturbations. However, the connection perturbations originating from PV term manifest in higher-order perturbation EOMs, such as the EOM of SIGWs. These modes of connection perturbations participate in nonlinear interactions without independent dynamics, which results in the strong coupling problem, as identified in Ref. \cite{Zhang:2023scq}. Introducing a non-minimally coupled boundary term to the action \eqref{action1} would incorporate connection perturbations into linear perturbation EOMs. This provides a viable pathway to resolve the strong coupling problem. This is the reason we consider the non-minimally coupled boundary term in this paper.

To avoid the aforementioned strong coupling problem, we consider the following action
\begin{equation}\label{action}
S_g=\int \mathrm{d}^4x\sqrt{-g}\left(-\frac{\mathring{R}}{2}-g_1(\varphi)B-g_2(\varphi) \widetilde{Q}Q\right)+\int \mathrm{d}^4x\sqrt{-g}\left(\frac{1}{2}g^{\mu\nu}\partial_{\mu}\varphi\partial_{\nu}\varphi-V(\varphi)\right),
\end{equation}
where $g_1$ and $g_2$ are coupling functions of the scalar field $\varphi$. Compared with the model in \cite{Li:2021mdp}, this action includes an additional non-minimally coupled boundary term. As shown in the following section, this term facilitates a resolution of the strong coupling problem.

Recent studies have investigated $f(\mathbb{Q},B)$ gravity in cosmology \cite{Bhoyar:2024goy,Subramaniam:2024uuu,Myrzakulov:2024jvg,Myrzakulov:2024pms,Shaily:2024tmx,Capozziello:2023vne,Capozziello:2024zij,Lohakare:2024oeu}, noting that $f(\mathring{R})$ gravity is recovered completely when $f(\mathbb{Q},B)=f(-\mathbb{Q}-B)$. The first two terms in action \eqref{action} can be viewed as a particular case of $f(\mathbb{Q},B)$ gravity featuring explicit non-minimal coupling to the boundary term.

\section{The evolution of background and linear perturbations}\label{sec3}
To compute the SIGWs in gravity model \eqref{action}, we must first establish the evolution of the background and linear scalar cosmological perturbations. In this section, we introduce the parameterization of both the metric and affine connection, and derive the EOMs for the background and the linear perturbations, from which we can see that the strong coupling problem is resolved effectively.

Consider the spatially flat FRW universe with small perturbations around it. In the Newtonian gauge, the metric takes the form
\begin{equation}\label{metric}
\mathrm{d}s^2=a^2\left\{(1+2\phi+2\phi^2)\mathrm{d}\tau^2-\left[(1-2\psi+2\psi^2)\delta_{ij}+h_{ij}+\frac{1}{2}h_{ik}h^{k}_{\ j}\right]\mathrm{d}x^i\mathrm{d}x^j\right\},
\end{equation}
where $\psi$, $\phi$ are linear scalar perturbations, and $h_{ij}$ is the second order tensor perturbation.

The connection components in symmetric teleparallel gravity are fully determined by four scalar fields $\xi^{\mu}$ Eq. \eqref{cy}, thus we can view the four scalar fields $\xi^{\mu}$ and metric $g_{\mu\nu}$ as the fundamental variables in symmetric teleparallel gravity. While the coincident gauge, $\xi^\mu=x^\mu$ significantly simplifies the cosmological perturbation calculation, this gauge may be incompatible with the parameterization of metric that we usually use in cosmology. Crucially, this coincident gauge does maintain compatibility with the spatially flat FRW universe at background level \cite{Zhao:2021zab}. We therefore adopt the background solution $\bar{\xi}^{\mu}=x^{\mu}$, and introduce perturbations $\delta\xi$ such that
\begin{equation}
\xi^{\mu}=\bar{\xi}^{\mu}+\delta \xi^{\mu}=x^{\mu}+\delta\xi^{\mu},
\end{equation}
 where $x^{\mu}$ are spacetime coordinates and $\delta \xi^{\mu}$ represents the small deviation from the background configuration, namely the perturbations of the scalar fields $\xi^{\mu}$. We further decompose $\delta\xi^{\mu}$ as $\delta\xi^{\mu}=\{C,\partial^{i}D\}$, where $C$ and $D$ are scalar perturbations. 
 
Similarly, the matter scalar field $\varphi$ is split as $\varphi + \delta\varphi$, where $\varphi(t)$ is the background value and $\delta\varphi$ represents the perturbation of the scalar field.

\subsection{The EOMs for the background and linear perturbations}
Expanding the action \eqref{action} to linear and quadratic order (the explicit expressions are given in appendix \ref{pactions}) and then varying with respect to the perturbations, we obtain the background EOMs
\begin{equation}\label{beom1}
\varphi''+2\mathcal{H}\varphi'+a^2V_\varphi+12g_{1\varphi}\mathcal{H}^2+6g_{1\varphi}\mathcal{H}'=0,
\end{equation}
\begin{equation}\label{beom2}
3\mathcal{H}^2-6\mathcal{H}g^{\prime}_{1}=\frac{1}{2}(\varphi')^2+a^2V,
\end{equation}
\begin{equation}\label{beom3}
-(\mathcal{H}^2+2\mathcal{H}')-2g^{'}_{1}\mathcal{H}+2g^{''}_{1}=\frac{1}{2}(\varphi')^2-a^2V,
\end{equation}
and the linear perturbation EOMs:
\begin{equation}\label{leom1}
\begin{split}
 & (6g_{1\varphi} \mathcal{H}+\varphi')\delta\varphi'+\left(6\mathcal{H}g'_{1\varphi}+a^2V_\varphi\right)\delta\varphi-2g_{1\varphi}\partial_i\partial^i\delta\varphi+6(\mathcal{H}-g^{'}_1)\psi' \\
 &- 2\partial_i\partial^i\psi+\left(6\mathcal{H}^2-\left(\varphi'\right)^2-12\mathcal{H}g^{'}_1\right)\phi+g^{'}_1\partial_i\partial^i\left(C-D'\right)=0,
\end{split}
\end{equation}
\begin{equation}\label{leom2}
\begin{split}
&6g_{1\varphi}\delta\varphi''-3\left(2g_{1\varphi}\mathcal{H}-4g^{'}_{1\varphi}+\varphi'\right)\delta\varphi'+
3\left(a^{2}V_{\varphi}-2\mathcal{H}g^{'}_{1\varphi}+2g^{''}_{1\varphi}\right)\delta\varphi-4g_{1\varphi}\partial_i\partial^{i}\delta\varphi \\ &+6\psi''+12\mathcal{H}\psi' - 9\left(2\mathcal{H}^{2}-\left(\varphi'\right)^{2}-8\mathcal{H}g^{'}_1-2\mathcal{H}'+2g^{''}_1\right)\psi-2\partial_i\partial^{i}\psi \\
 &+6\left(\mathcal{H}-g^{'}_1\right)\phi'+ 6\left(2\mathcal{H}^{2}-2\mathcal{H}g^{'}_1+\mathcal{H}'-g^{''}_1\right)\phi+2\partial_i\partial^{i}\phi+g^{'}_1\partial_i\partial^{i}\left(C+3D'\right)=0,
\end{split}
\end{equation}
\begin{equation}\label{leom3}
\begin{split}
&\delta\varphi''+2\mathcal{H}\delta\varphi'+\left(a^{2}V_{\varphi\varphi}+12g_{1\varphi\varphi}\mathcal{H}^{2}+6g_{1\varphi\varphi}\mathcal{H}'\right)\delta\varphi-\partial_i\partial^i\delta\varphi \\&-6g_{1\varphi}\psi''-3\left(10g_{1\varphi}\mathcal{H}+\varphi'\right)\psi' +4g_{1\varphi}\partial_i\partial^i\psi-\left(6g_{1\varphi}\mathcal{H}+\varphi'\right)\phi'\\&-2\left(12g_{1\varphi}\mathcal{H}^{2}+6g_{1\varphi}\mathcal{H}'+2\mathcal{H}\varphi'+\varphi''\right)\phi-2g_{1\varphi}\partial_i\partial^i\phi-2g_{1\varphi}\mathcal{H}\partial_i\partial^i\left(C+D'\right)=0,
\end{split}
\end{equation}
\begin{equation}\label{leom4}
\begin{split}
&2g_{1\varphi}\mathcal{H}\partial_i\partial^i\delta\varphi-g^{'}_1 \partial_i\partial^i\left(\phi-\psi\right)-(2g^{'}_1\mathcal{H}+g^{''}_1)\partial_i\partial^i\left(C-D'\right) -g^{'}_1\partial_i\partial^i\left(\partial_j\partial^j D-D''\right)=0,
\end{split}
\end{equation}
\begin{equation}\label{leom5}
\begin{split}
&2g_{1\varphi}\mathcal{H}\partial_{i}\partial^{i}\delta\varphi'+2\left(2g_{1\varphi}\mathcal{H}^{2}+\mathcal{H}g^{'}_{1\varphi}+g_{1\varphi}\mathcal{H}'\right)\partial_{i}\partial^{i}\delta\varphi+3g^{'}_1\partial_{i}\partial^{i}\psi' \\& +3(2\mathcal{H}g^{'}_1+g^{''}_1)\partial_{i}\partial^{i}\psi+g^{'}_1\partial_{i}\partial^{i}\phi'+(g^{''}_1+2\mathcal{H}g^{'}_1)\partial_{i}\partial^{i}\phi- g^{'}_1\partial_{i}\partial^{i}C'' \\&+g^{'}_1\partial_{j}\partial^{j}\partial_{i}\partial^{i}C -(2\mathcal{H}g^{'}_1+g^{''}_1)\partial_{i}\partial^{i}(C'-\partial_{j}\partial^{j}D)=0,
\end{split}
\end{equation}
where primes denote derivatives with respect to the conformal time, $g_{1\varphi}=\text{d}g_1/\text{d}\varphi$, $g_{1\varphi\varphi}=\text{d}^2 g_1/\text{d} \varphi^2$, $V_\varphi=\text{d}V/\text{d}\varphi$, and $V_{\varphi\varphi}=\text{d}^2 V/\text{d} \varphi^2$.

An analysis of the EOMs \eqref{beom1}-\eqref{beom3} and \eqref{leom1}-\eqref{leom5} reveals that a constant coupling function $g_1(\varphi)$
eliminates the boundary term's contribution to the system, reducing the background equations to those of GR. In this case, the connection perturbations $C$ and $D$ disappear entirely from the linear EOMs. However, as demonstrated in the following section, $C$ and $D$ reappear in the EOMs of SIGWs through the PV term, which leads to the strong coupling problem. By introducing the non-minimally coupled boundary term, this strong coupling problem is effectively resolved provided $g_1(\varphi)$ is not constant.

\subsection{The solutions for background and linear perturbations}
Solving the complex background equations \eqref{beom1}-\eqref{beom3} and linear perturbation equations \eqref{leom1}-\eqref{leom5} analytically is difficult for arbitrary coupling function $g_1(\varphi)$ and cosmological epochs. In this paper, we consider the SIGWs generated during the radiation-dominated era. In this epoch, the universe expands according to a power law $a=a_i\tau/\tau_i$. Furthermore, we adopt the linear coupling function $g_1(\varphi)=c\varphi$, which preserves the shift symmetry of the boundary term and simplifies the subsequent calculations.

The equations yield the solutions for background
\begin{equation}\label{aH}
\mathcal{H}=\frac{1}{\tau},
\end{equation}
and
\begin{equation}
\varphi(\tau)=-2 c \log \left(\tau^{\sqrt{4+25c^2}/c}+C_1\right)+\left(\sqrt{25 c^2+4}-5 c\right) \log (\tau)+C_2,
\end{equation}
where $C_1$ and $C_2$ are integration constants. If we choose $C_1=0$, then $\varphi(\tau)\propto\log (\tau)$, which coincides with the GR solution \cite{Zhang:2023scq}. Consequently, the scalar field reduces to
\begin{equation}\label{solvarphi}
\varphi(\tau)=\mathbb{C}\log (\tau/\tau_0),
\end{equation}
with the coefficient $\mathbb{C}=-\left(5 c+\sqrt{25 c^2+4}\right)$, and $\tau_0=\text{e}^{-C_2/\mathbb{C}}$.

To facilitate solving the linear EOMs and the subsequent calculation of SIGWs, we perform a Fourier transform on the EOM set Eqs. \eqref{leom1}-\eqref{leom5} and decompose the scalar perturbations into the primordial components and the transfer functions as follows
\begin{equation}
\delta\varphi(\bm k, \tau)=\frac{2}{3}\zeta(\bm k) T_{\delta\varphi}(x),
\end{equation}
\begin{equation}
\phi(\bm k, \tau)=\frac{2}{3}\zeta(\bm k) T_{\phi}(x),
\end{equation}
\begin{equation}
\psi(\bm k, \tau)=\frac{2}{3}\zeta(\bm k) T_{\psi}(x),
\end{equation}
\begin{equation}
C(\bm k, \tau)=\frac{2}{3}\zeta(\bm k)\frac{1}{k}T_{C}(x),
\end{equation}
and
\begin{equation}
D(\bm k, \tau)=\frac{2}{3}\zeta(\bm k)\frac{1}{k^2}T_{D}(x).
\end{equation}
where $x=k\tau$ and $\zeta$ denotes the primordial perturbation. Consequently, the linear perturbation equations \eqref{leom1}-\eqref{leom5} can be reformulated in terms of transfer functions as follows
\begin{equation}\label{leom12}
\begin{split}
 & (6c+\mathbb{C})xT^{*}_{\delta\varphi}(x)-(6 c+\mathbb{C})T_{\delta\varphi}(x)+2c x^2 T_{\delta\varphi}(x)+6(1-c \mathbb{C})x T^{*}_{\psi}(x) \\
 &+2x^2T_{\psi}(x)-(\mathbb{C}^2+12 c \mathbb{C}-6)T_{\phi}(x)-c\mathbb{C}x\left(T_{C}(x)-T^{*}_D(x)\right)=0,
\end{split}
\end{equation}
\begin{equation}\label{leom22}
\begin{split}
&6c x^2T^{**}_{\delta\varphi}(x)-3(2c+\mathbb{C})xT^{*}_{\delta\varphi}(x)-
3(6 c+\mathbb{C})T_{\delta\varphi}(x)+4c x^2T_{\delta\varphi}(x)+6x^2T^{**}_{\psi}(x)\\&+12x T^{*}_{\psi}(x)-9(4-\mathbb{C}^2-10c\mathbb{C})T_{\psi}(x)+2x^2T_{\psi}(x) +6(1-c \mathbb{C})x T^{*}_{\phi}(x)\\&+ 6(1-c \mathbb{C})T_{\phi}(x)-2x^2T_{\phi}(x)-c \mathbb{C}x\left(T_{C}(x)+3T^{*}_{D}(x)\right)=0,
\end{split}
\end{equation}
\begin{equation}\label{leom32}
\begin{split}
&\mathbb{C}x^2T^{**}_{\delta\varphi}+2\mathbb{C}x T^{*}_{\delta\varphi}(x)-4(6c+\mathbb{C})T_{\delta\varphi}(x)+\mathbb{C}x^2T_{\delta\varphi}(x) -6c\mathbb{C} x^2 T^{**}_{\psi}\\&-3(10 c+\mathbb{C})\mathbb{C}x T^{*}_{\psi} -4c\mathbb{C} x^2T_{\psi}(x)-(6 c+\mathbb{C})\mathbb{C}x T^{*}_{\phi}(x)-2(6 c+\mathbb{C})\mathbb{C}T_{\phi}(x)\\&+2c\mathbb{C} x^2T_{\phi}(x)+2c\mathbb{C} x\left(T_C(x)+T^{*}_{D}(x)\right)=0,
\end{split}
\end{equation}
\begin{equation}\label{leom42}
\begin{split}
&2xT_{\delta\varphi}(x)-\mathbb{C}x\left(T_{\phi}(x)-T_{\psi}(x)\right)-\mathbb{C}\left(T_C(x)-T^{*}_D(x)\right) +\mathbb{C}x\left(T_D(x)+T^{**}_{D}(x)\right)=0,
\end{split}
\end{equation}
\begin{equation}\label{leom52}
\begin{split}
&2xT^{*}_{\delta\varphi}+2T_{\delta\varphi}(x)+3\mathbb{C}xT^{*}_{\psi}(x) +3\mathbb{C}T_{\psi}(x)+\mathbb{C}x T^{*}_{\phi}(x)+\mathbb{C}T_{\phi}(x)-\mathbb{C}x T^{**}_{C}(x) \\&-\mathbb{C}xT_{C}(x) -\mathbb{C}(T^{*}_C(x)+T_{D}(x))=0,
\end{split}
\end{equation}
where ``$*$" represents the derivative with respect to the argument.

The complexity of the perturbation EOMs \eqref{leom12}-\eqref{leom52} necessitates numerical solutions. Moreover, diffeomorphism invariance of the action \eqref{action} implies that not all of these equations are independent \cite{BeltranJimenez:2018vdo,Jarv:2018bgs,Hohmann:2021fpr,Li:2021mdp}, further complicating direct numerical treatment. In GR, by taking the scalar part of the off-diagonal $ij$ component of the Einstein equation, we can obtain $\phi=\psi$ when anisotropic stress is negligible. Applying the same method to \eqref{action}, we get the relation between $\phi$ and $\psi$,
\begin{equation}\label{leom6}
\phi-\psi+2g^{'}_{1}C+2g_{1\varphi}\delta\varphi=0,
\end{equation}
which, writing the above equation in Fourier space and in terms of transfer functions, becomes
\begin{equation}\label{leom62}
xT_\phi(x)-xT_\psi(x)+2c\mathbb{C}T_C(x)+2cxT_{\delta\varphi}(x)=0.
\end{equation}
We will solve the linear perturbation EOMs including Eq. \eqref{leom62} numerically and subsequently compute the SIGWs.

\section{The SIGWs}\label{sec4}
Expanding the action \eqref{action} to cubic order, we obtain the action of SIGWs \eqref{acgws}. From the cubic action, the perturbations $C$ and $D$ appear in the action of the SIGWs even without including the boundary term. The presence of the non-minimally coupled boundary term resolves the strong coupling problem.

Varying the action of SIGWs with respect to $h^{ij}$ yields the EOM for SIGWs. In Fourier space, it takes the form
\begin{equation}\label{eu1}
u^{A''}_{\bm k}+\left(\omega^2_A-\frac{a''}{a}\right)u^{A}_{\bm k}=4aS^A_{\bm{k}},
\end{equation}
where  $u^A_{\bm k}=ah^A_{\bm k}$, $h^A_{\bm k}$ denotes the polarized tensor modes, and
\begin{equation}
\omega^2_A=k^2-4\mathcal{M}\lambda^Ak, \ \ (\lambda^R=1,\ \lambda^L=-1),
\end{equation}
the definitions of $\mathcal{M}$ and the source term $S^A_{\bm{k}}$ are given explicitly in the appendix. 

The EOMs for SIGWs \eqref{eu1} can be solved using the Green’s function method. Under the exponential coupling that ensures a time‑independent GW speed \cite{Zhang:2023scq,Zhang:2024vfw}, the analytic Green’s function takes the form given in Eq. \eqref{greenfunction}. After a lengthy but straightforward calculation (The detailed calculation procedure can be found in the appendix \ref{calculationgws}), we obtain the power spectrum of SIGWs as follows
\begin{align}
\label{PStensor}
\mathcal{P}^{A}_h(k,x)=4\int_{0}^\infty\mathrm{d}u\int_{|1-u|}^{1+u}\mathrm{d}v
\mathcal{J}(u,v)I^{A}(u,v,x)^2\mathcal{P}_\zeta(uk)\mathcal{P}_\zeta(vk),
\end{align}
where
\begin{equation}
\mathcal{J}(u,v)=\left[\frac{4u^2-(1+u^2-v^2)^2}{4uv}\right]^2,
\end{equation}
and $\mathcal{P}_{\zeta}$ is the power spectrum of the primordial curvature perturbation.

The fractional energy density of the SIGWs is given by \cite{Maggiore:1999vm,Cai:2019jah,Kohri:2018awv,Espinosa:2018eve} \footnote{In this paper, we use the parametrized metric \eqref{metric} without the additional factor $1/2$ in front of the tensor perturbation $h_{ij}$, which differs from conventions adopted in some parts of the literature. Consequently, the prefactor in the first line of Eq. \eqref{OGW1} is $1/12$ instead of $1/48$. The final results are independent of the metric parametrization, as stated in \cite{Garcia-Saenz:2023zue}.}
\begin{equation}\label{OGW1}
\begin{split}
\Omega_{\mathrm{GW}}(k,x)&=\frac{1}{12}\left(\frac{k}{\mathcal{H}}\right)^2\sum\limits_{A=R,L}\overline{\mathcal{P}^A_h(k,x)}=\frac{x^2}{12}\sum\limits_{A=R,L}\overline{\mathcal{P}^A_h(k,x)}\\
&=\frac{1}{3}\int_{0}^\infty\mathrm{d}u\int_{|1-u|}^{1+u}\mathrm{d}v
\mathcal{J}(u,v)\sum\limits_{A=R,L}\overline{\tilde{I}^{A}(k,u,v,x)^2}\mathcal{P}_\zeta(uk)\mathcal{P}_\zeta(vk),
\end{split}
\end{equation}
where $\overline{\tilde{I}^{A}(k,u,v,x)^2}=\overline{I^{A}(k,u,v,x)^2}x^2$, their definitions are provided in appendix \ref{calculationgws}.

Since the GWs behave as free radiation, and taking the thermal history of the universe into account, the energy density spectrum of the SIGWs at the present time $\Omega_{\mathrm{GW},0}$ is \cite{Espinosa:2018eve}
\begin{equation}\label{EGW}
\Omega_{\mathrm{GW},0}\left(k\right)=\Omega_{\mathrm{GW}}\left(k,\eta\rightarrow\infty\right)\Omega_{r,0},
\end{equation}
where $\Omega_{r,0} \approx 9\times 10^{-5}$ represents the current fractional energy density of the radiation \cite{Sato-Polito:2019hws}.

The absence of analytic solutions necessitates a numerical approach, both for solving the scalar perturbation EOMs and for computing the SIGWs. For simplicity, we adopt a monochromatic primordial curvature power spectrum to compute the SIGW energy density spectrum, thereby analyzing the SIGW features predicted by our model in \eqref{action},
\begin{equation}\label{ps1}
\mathcal{P}_\zeta(k)=\mathcal{A}_\zeta\delta(\ln(k/k_p)),
\end{equation}
which yields the present-day energy density spectrum \footnote{
As the analysis in appendix \ref{calculationgws}, observational constraints imply that the PV term is suppressed by $\mathcal{M}/k$ and thus negligible; we therefore omit its contribution.}
\begin{align}
\label{ogw}
\Omega_{\text{GW},0}(k)=\frac{1}{3}\Omega_{r,0}\mathcal{A}_{\zeta}^2\tilde{k}^{-2}\mathcal{J}(\tilde{k}^{-1},\tilde{k}^{-1})\sum\limits_{A=R,L}\overline{\tilde{I}^{A}(k,\tilde{k}^{-1},\tilde{k}^{-1},x\rightarrow \infty)^2}\Theta(2-\tilde{k}),
\end{align}
where $\tilde{k}=k/k_p$. 

We compute the energy density of SIGWs numerically, the results are shown in Fig. \ref{fig:GWs}. Although the PV term is negligible, the effect of the boundary term on  SIGWs is significant, particularly in the high-frequency regime. The existence of non-minimally coupled boundary term not only helps to avoid the strong coupling problem, but also alters the evolution of the scalar perturbations, leading to distinct SIGW signatures compared to GR. At low frequencies, the behavior of SIGWs is similar to that in GR. However, at high frequencies, they exhibit different features. For the SIGWs in GR, there is a divergence at $\tilde{k}=2/\sqrt{3}$ due to the resonant amplification \cite{Ananda:2006af,Kohri:2018awv} \footnote{This divergence arises only for a monochromatic spectrum. For a finite-width narrow spectrum, the resonant divergence is smoothed out, as more modes contribute to the SIGWs.}. In our model, the spectrum of energy density of SIGWs is regular across all frequencies, and the peaks in SIGWs exhibit an obvious rightward shift relative to those in GR. These distinctive features can be detected by the future space-based GW detectors, such as LISA, TianQin and Taiji \cite{Robson:2018ifk,Luo:2015ght,Gong:2021gvw,Ruan:2018tsw}.

\begin{figure}[htp]
\centering
\includegraphics[width=0.75\linewidth]{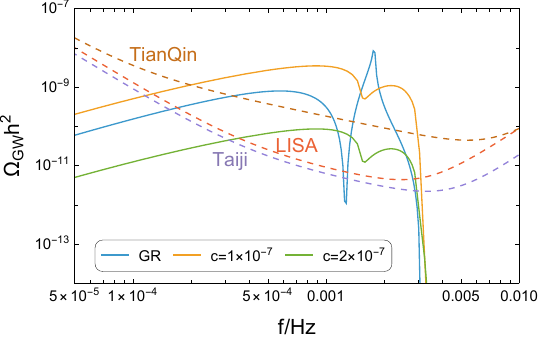}
\caption{The energy density spectrum of SIGWs generated by the monochromatic primordial power spectrum with the amplitude $\mathcal{A}_{\zeta}=10^{-2}$. The peak scale $k_p=10^{12} \text{Mpc}^{-1}$ corresponds to the maximum sensitivity range of GW detectors. The dashed curves represent the sensitivity curves of LISA, TianQin and Taiji, while the solid curves represent the energy density spectrum of SIGWs in GR and PV symmetric teleparallel gravity, respectively.}
\label{fig:GWs}
\end{figure}

\section{Conclusion}\label{seccon}
The phenomenon of parity violation in gravity has recently attracted significant theoretical interest. In Ref. \cite{Li:2021mdp}, the authors proposed a simple PV gravity model within the framework of symmetric teleparallel gravity, and further studied the linear cosmological perturbations. To resolve the strong coupling problem inherent in this model, we introduced a non-minimally coupled boundary term described by action \eqref{action}. This non-minimally coupled boundary term incorporates the connection perturbations $C$ and $D$ into the linear scalar perturbation EOMs \eqref{leom1}-\eqref{leom5}, thereby eliminating the strong coupling problem, provided that the coupling function $g_1(\varphi)$ is not constant.

After resolving the strong coupling problem, we derived the EOMs for linear perturbations and for the SIGWs. We then calculated the SIGWs in model \eqref{action} during the  radiation-dominated era. To maintain the shift symmetry of the boundary term while ensuring computational tractability, we chose the coupling function $g_{1}(\varphi)=c\varphi$. For the PV term, we adopt an exponential coupling form that guarantees time-independent SIGW propagation speed. By solving the scalar perturbation equations numerically, we calculated the SIGW energy density spectrum with the monochromatic power spectrum \eqref{ps1}. The non-minimally coupled boundary term not only eliminates the strong coupling problem but also modifies scalar perturbation evolution, leading to distinct SIGW signatures, particularly at high frequencies compared to GR. The peaks in the energy density spectrum of SIGWs exhibit an obvious rightward shift relative to GR predictions, this distinctive feature can be detected by future space-based GW detectors, such as LISA, TianQin and Taiji.

\appendix
\section{The perturbed actions}\label{pactions}
In this appendix, we present the perturbed actions. Expanding the gravity model \eqref{action} to linear and quadratic order, we obtain the linear action,
\begin{equation}\label{linearaction}
\begin{split}
S_1=&\int \text{d}x^3 \text{d}\tau a^2\left[-\left(a^2V_\varphi+2\mathcal{H}\varphi'+\varphi''+12g_{1\varphi}\mathcal{H}^2+6g_{1\varphi}\mathcal{H}'\right)\delta\varphi\right.\\&\left.-\left(\frac{1}{2}(\varphi')^2+a^2V-3\mathcal{H}^2+6g^{'}_{1}\mathcal{H}\right)\phi \right.\\&\left.-3\left(\frac{1}{2}(\varphi')^2-a^2V+2\mathcal{H}'+\mathcal{H}^2+2g^{'}_{1}\mathcal{H}-2g^{''}_{1}\right)\psi\right],    
\end{split}
\end{equation}
and the quadratic action
\begin{equation}\label{S2}
\begin{split}
S_2=&\int \text{d}x^3\text{d}\tau a^2\left[-\frac{1}{2}\partial_i\delta\varphi\partial^{i}\delta\varphi-a^{2}V_{\varphi}(\phi-3\psi)\delta\varphi-\frac{1}{2}a^{2}V_{\varphi\varphi}\delta\varphi^{2}+\frac{1}{2}(\delta\varphi')^{2}\right.\\&\left.-
(\phi+3\psi)\delta\varphi'\varphi'+(2\phi-\psi)\partial^i \partial_i \psi \right.\\& \left.-
3\left(2\mathcal{H}\phi\psi'+6\mathcal{H}\psi\psi'+\left(\psi'\right)^2-6g^{'}_1\psi\psi'-2g^{'}_1\phi\psi'\right) \right.\\&\left.-
\left(9\psi^2+\phi^2\right)\left(3\mathcal{H}^2-\frac{1}{2}\left(\varphi'\right)^2-6g^{'}_1\mathcal{H}\right)-3g_{1\varphi\varphi}\left(\mathcal{H}'+2\mathcal{H}^2\right)\delta\varphi^2 \right.\\& \left.+
6g_{1\varphi}\left(\phi\mathcal{H}'+\mathcal{H}\phi'+2\mathcal{H}^2\phi+3\psi\mathcal{H}'+5\mathcal{H}\psi'
+\psi''+6\mathcal{H}^2\psi\right)\delta\varphi \right.\\&\left.+
2g_{1\varphi}\left(\partial_i\partial^i\phi-2\partial_i\partial^i\psi\right)\delta\varphi+2g_{1\varphi} \mathcal{H}\delta\varphi\left(\partial^i\partial_iD'+\partial_i\partial^iC\right) \right.\\&\left.-
g^{'}_1\phi\left(\partial^i\partial_iC-\partial^i\partial_iD'\right)+g^{'}_1\psi\left(3\partial^i\partial_iD'+\partial^i\partial_iC\right) \right.\\&\left.+
g^{'}_1\left(C'-\partial^j\partial_jD\right)\left(\partial^i\partial_iC-\partial^i\partial_iD'\right)\right].
\end{split}
\end{equation}
Varying the above actions yields the EOMs for the background and linear perturbations.

Expanding action \eqref{action} to cubic order gives the action for SIGWs
\begin{equation}\label{acgws}
S_{\text{GW}}=S_{TT}+S_{SST},
\end{equation}
where the quadratic action for the tensor perturbation is
\begin{equation}
S_{TT}=\int \mathrm{d}^3x \mathrm{d}\tau a^2\left[\frac{1}{8}\left(h^{'}_{ij}h^{'ij}-\partial_k h_{ij}\partial^k h^{ij}\right)+\frac{1}{2}\mathcal{M}\epsilon^{ijk}\partial_j h_{kl}h^{\ l}_{i}\right],
\end{equation}
with
\begin{equation}\label{MM}
\mathcal{M}=2(2\mathcal{H}g_2(\varphi)+g^{'}_2(\varphi)).
\end{equation}

The cubic scalar-scalar-tensor interaction takes the form
\begin{equation}
S_{SST}=\int \text{d}x^3 \text{d}\tau a^2\left(\mathcal{L}^{1}_{ij}+\mathcal{L}^{B}_{ij}+\mathcal{L}^{\text{PV}}_{ij}\right)h^{ij},
\end{equation}
where
\begin{equation}\label{L1}
\begin{split}
\mathcal{L}^{1}_{ij}= &\frac{1}{2}\partial_i\delta\varphi\partial_j\delta\varphi+\partial_i\phi\partial_j\psi,
\end{split}
\end{equation}
$\mathcal{L}^{B}_{ij}$ is the contribution from the boundary term,
\begin{equation}\label{L2}
\begin{split}
\mathcal{L}^{B}_{ij}=&-2g_{1\varphi}\partial_{i}\partial_{j}(\phi-\psi)\delta\varphi +g^{'}_1(\phi+\psi)\partial_{i}\partial_{j}C+g^{'}_{1\varphi}\delta\varphi \partial_{i}\partial_{j}C \\& +g_{1\varphi}(\partial^{k}\delta\varphi \partial_{k}\partial_{i}\partial_{j}D+\delta\varphi \partial^{k}\partial_{k}\partial_{i}\partial_{j}D+\delta\varphi' \partial_{i}\partial_{j}C+\delta\varphi \partial_{i}\partial_{j}C') \\& -g^{'}_1\left(C' \partial_{i}\partial_{j}C-\partial^{k}C \partial_{k}\partial_{i}\partial_{j}D\right),
\end{split}
\end{equation}
and 
\begin{equation}\label{LPV}
\begin{split}
\mathcal{L}^{\mathrm{PV}}_{ij}= &\mathcal{M}\epsilon_{jkl}(\partial^k\partial^m D\partial^l\partial_m\partial_i D-\partial^k\partial_i C\partial^l D')+2g_{2\varphi}\epsilon_{jkl}\partial^l\delta\varphi(\partial^k\partial_i C+\partial^k\partial_i D')\\&+2g_2\epsilon_{jkl}(2\partial^k\partial_i C\partial^l\phi-2\partial^k\partial_i D'\partial^l\phi+4\partial^k\psi\partial^l\partial_i D'\\&-\partial^k\partial_i C\partial^l C'-\partial^k C'\partial^l\partial_i D'-\partial^k\partial_i C\partial^l D''-2\partial^k\partial_i\partial_m D\partial^l\partial^m C\\&+2\partial^k\partial^m D\partial^l\partial_m\partial_i D'-\partial^k D''\partial^l\partial_i D'),
\end{split}
\end{equation}
which corresponds to the PV term's contribution, and $g_{2\varphi}=\text{d}g_2/\text{d}\varphi$.

\section{The calculation of the SIGWs}\label{calculationgws}
Varying the action \eqref{acgws} with respect to $h^{ij}$ yields the EOM for SIGWs,
\begin{equation}\label{EOMGWs}
h^{''}_{ij}+2\mathcal{H}h^{'}_{ij}-\nabla^2h_{ij}-2\mathcal{M}\left(\epsilon_{ilk}\partial_l h_{kj}+\epsilon_{jlk}\partial_l h_{ki}\right)=4\mathcal{T}^{lm}_{\ \ ij}s_{lm},
\end{equation}
where $\mathcal{T}^{lm}_{\ \ ij}$ is the projection tensor that is used to extract the transverse and trace-free part of the source,
\begin{equation}\label{TT}
\mathcal{T}^{lm}_{\ \ \ ij}s_{lm}(\bm{x},\tau)=\sum\limits_{A=R,L}\int \frac{\mathrm{d}^3\bm k}{(2\pi)^{3/2}}e^{i {\bm k} \cdot {\bm x}}p_{ij}^A p^{Alm}\tilde{s}_{lm}(\bm k,\tau), 
\end{equation}
with $\tilde{s}_{ij}$ denoting the Fourier transform of the source $s_{ij}$,
\begin{equation}
s_{ij}=\frac{1}{2}(\mathcal{L}^{1}_{ij}+\mathcal{L}^{B}_{ij}+\mathcal{L}^{\mathrm{PV}}_{ij}+i \leftrightarrow j).
\end{equation}

In Eq. \eqref{TT}, $p_{ij}^A$ are the circular polarization tensors defined as
\begin{equation}\label{cpt}
p^R_{ij}=\frac{1}{\sqrt{2}}(\mathbf e^{+}_{ij}+i\mathbf e^{\times}_{ij}), \ \ p^L_{ij}=\frac{1}{\sqrt{2}}(\mathbf e^{+}_{ij}-i\mathbf e^{\times}_{ij}),
\end{equation}
where $\mathbf e^+_{ij}$ and $\mathbf e^{\times}_{ij}$ represent the plus and cross polarization tensors, respectively, and can be expressed as
\begin{equation}
\label{poltensor1}
\begin{split}
\mathbf e^+_{ij}=&\frac{1}{\sqrt{2}}(\mathbf e_i \mathbf e_j-\bar{\mathbf e}_i \bar{\mathbf e}_j),\\
\mathbf e_{ij}^\times=&\frac{1}{\sqrt{2}}(\mathbf e_i\bar{\mathbf e}_j+\bar{\mathbf e}_i \mathbf e_j),
\end{split}
\end{equation}
where ${\mathbf e_{i}\left(\bm{k}\right)}$ and ${\bar{\mathbf e}_{i}\left(\bm{k}\right)}$ form an orthonormal basis orthogonal to the wavevector ${\bm{k}}$.

We can rewrite the EOM for SIGWs in Fourier space as
\begin{equation}\label{eu}
u^{A''}_{\bm k}+\left(\omega^2_A-\frac{a''}{a}\right)u^{A}_{\bm k}=4aS^A_{\bm{k}},
\end{equation}
where  $u^A_{\bm k}=ah^A_{\bm k}$, and $h^A_{\bm k}$ represent the polarized modes
\begin{equation}
h_{ij}(\bm{x},\tau)=\sum\limits_{A=R,L}\int \frac{\mathrm{d}^3k}{(2\pi)^{3/2}}e^{i\bm{k}\cdot\bm{x}}p^{A}_{ij}h^A_{\bm{k}}(\tau).
\end{equation}
In Eq. \eqref{eu}, the angular frequency is defined as
\begin{equation}\label{omgA}
\omega^2_A=k^2-4\mathcal{M}\lambda^Ak, \ \ (\lambda^R=1,\ \lambda^L=-1),
\end{equation}
with $\mathcal{M}$ given in Eq. \eqref{MM}, and the source
\begin{equation}
S^A_{\bm{k}}=p^{Aij}\tilde{s}_{ij}(\bm{k},\tau).
\end{equation}

The source term $S^A_{\bm k}$ contains two parts: the parity-conserving part and the PV part:
\begin{equation}
S^A_{\bm k}=S^{A(\mathrm{PC})}_{\bm k}+S^{A(\mathrm{PV})}_{\bm k},
\end{equation}
where 
\begin{equation}
\begin{split}
 S^{A(\mathrm{PC})}_{\bm k}=&\int \frac{\mathrm{d}^3\bm k'}{(2\pi)^{3/2}}p^{Aij}k^{'}_i k^{'}_j\zeta(\bm k')\zeta(\bm k-\bm k') f_{\mathrm{PC}}(u,v,x),\\
 S^{A(\mathrm{PV})}_{\bm k}=&\int \frac{\mathrm{d}^3\bm k'}{(2\pi)^{3/2}}p^{Aij}k^{'}_i k^{'}_j\zeta(\bm k')\zeta(\bm k-\bm k') f_{\mathrm{PV}}(k,u,v,x),
\end{split}
\end{equation}
with  $u=k'/k$, $v=|\bm k-\bm{k}'|/k$, and
\begin{equation}
p^{Aij}k^{'}_i k^{'}_j=\frac{1}{2}k^{'2}\sin^2(\theta)\text{e}^{2i\lambda^A\ell},
\end{equation}
where $\theta$ is the angle between $\bm{k}'$ and $\bm{k}$, while $\ell$ is the azimuthal angle of $\bm{k}'$. The parity-conserving part $f_{\mathrm{PC}}(u,v,x)=f_{\mathrm{PC1}}(u,v,x)+f_{\mathrm{PC2}}(u,v,x)$ contains two components:
\begin{equation}\label{fpc1}
\begin{split}
f_{\mathrm{PC1}}(u,v,x)=&\frac{2}{9}\left[\frac{1}{2}T_{\delta\varphi}(ux)T_{\delta\varphi}(vx)+T_{\phi}(ux)T_{\psi}(vx)+(u\leftrightarrow v)\right],
\end{split}
\end{equation}
and $f_{\mathrm{PC2}}(u,v,x)$ represents the boundary term's contribution,
\begin{equation}\label{fpc2}
\begin{split}
f_{\mathrm{PC2}}(u,v,x)=&\frac{2}{9}c\left[\frac{1-u^2+v^2}{2v^2}T_{\delta\varphi}(ux)T_{D}(vx)-\frac{u}{v}T^{*}_{\delta\varphi}(ux)T_{C}(vx)\right.\\&\left.-T_{\delta\varphi}(ux)T^{*}_{C}(vx)+(T_{\phi}(ux)-T_{\psi}(ux))T_{\delta\varphi}(vx)\right.\\&\left.+\mathbb{C}\left(\frac{T^{*}_C(ux)T_C(vx)}{vx}+\frac{1-u^2-v^2}{2uv^2x}T_C(ux)T_{D}(vx)\right)\right.\\&\left.-\mathbb{C}\frac{T_{\phi}(ux)+T_{\psi}(ux)}{vx}T_C(vx)+(u\leftrightarrow v)\right].
\end{split}
\end{equation}

$f^A_{\mathrm{PV}}(u,v,x)$ corresponds to the contribution from the PV term,  
\begin{equation}\label{fPV}
\begin{split}
f^A_{\mathrm{PV}}(u,v,x)= &\frac{2}{9}\lambda^A\left[\frac{\mathcal{M}}{k}\left(\frac{1-u^2-v^2}{2uv^2}T_D(ux)T_D(vx)-\frac{1}{v}T_C(ux)T^{*}_D(vx)\right)\right.\\&\left.+2g_{2\varphi}\frac{u}{v}T_{\delta\varphi}(ux)(T_C(vx)+T^{*}_D(vx))\right.\\&\left.+2g_2\left(2T_{\phi}(ux)(T_C(vx)-T^{*}_D(vx))-\frac{4u}{v}T_{\psi}(ux)T^{*}_{D}(vx)\right.\right.\\&\left.\left.+\frac{v}{u}T_{C}(ux)(T^{*}_C(vx)+T^{**}_D(vx))+\frac{u}{v}T^{*}_C(ux)T^{*}_D(vx)\right.\right.\\&\left.\left.+2\frac{1-u^2-v^2}{uv}T_D(ux)T_{C}(vx)+2\frac{1-u^2-v^2}{uv}T_D(ux)T^{*}_{D}(vx)\right.\right.\\&\left.\left.+\frac{u}{v}T^{**}_D(ux)T^{*}_D(vx)\right)+u\leftrightarrow v\right].
\end{split}
\end{equation}

Eq. \eqref{eu} can be solved by the method of Green's function,
\begin{equation}\label{Eh}
h^{A}_{\bm k}\left(\tau\right)=\frac{4}{a(\tau)}\int^{\tau}_{0}\mathrm{d}\bar{\tau}~G^A_{k}\left(\tau,\bar{\tau}\right)
a\left(\bar{\tau}\right)S^A_{\bm k}\left(\bar{\tau}\right),
\end{equation}
where the Green's function $G^A_{k}\left(\tau,\bar{\tau}\right)$ satisfies the following differential equation:
\begin{equation}\label{GREEN}
G^{A''}_{k}(\tau,\bar{\tau})+\left(\omega_A^2-\frac{a''}{a}\right)G^{A}_{k}(\tau,\bar{\tau})=\delta(\tau-\bar{\tau}).
\end{equation}

For a general coupling function $g_{2}(\varphi)$, the angular frequency $\omega_A$ in the above equation depends complexly on both the conformal time $\tau$ and the wavenumber $k$, making analytic solution of Eq. \eqref{GREEN} intractable. Moreover, $\omega_A$ is associated with the propagation speed of the GWs, therefore, we assume that $\omega_A$ is approximately time-independent during the generation and propagation of SIGWs. 

As argued in \cite{Zhang:2023scq,Zhang:2024vfw}, the exponential coupling function
$g_2(\varphi)=g_0\mathrm{e}^{\alpha\varphi}$ renders $\mathcal{M}=\mathcal{M}_0=6g_0/\tau_0$ and hence $\omega_A$ time-independent under the condition $\alpha\mathbb{C}-1=0$. This choice allows an analytic solution of the Green’s function equation
\begin{equation}\label{greenfunction}
G^{A}_{k}(\tau,\bar\tau)=\frac{\sin[\omega_A(\tau-\bar\tau)]}{\omega_A}\Theta(\tau-\bar\tau),
\end{equation}
where $\Theta$ is the Heaviside step function.

Considering the observational constraints on the speed of GWs from joint analyses of GW and gamma‑ray burst events, which indicate that the deviation of the GW speed from the speed of light is very small \cite{TheLIGOScientific:2017qsa,LIGOScientific:2017zic}, we conclude that $\mathcal{M}_0/k \ll 1$. Furthermore, LIGO‑Virgo GW events also constrain $\mathcal{M}_0/k \ll 1$ \cite{Wu:2021ndf}. Recalling the EOM for SIGWs \eqref{EOMGWs} and the source term \eqref{fPV}, the contribution from the PV term is suppressed by $|\mathcal{M}_0|/k$, implying that the effect of the PV term on SIGWs is negligible.

For convenience, we rewrite Eq. \eqref{Eh} as
\begin{equation}
\label{hsolution}
h^A_{\bm k}(\tau)=4 \int\frac{\mathrm{d}^3\bm k'}{(2\pi)^{3/2}} p^{Aij}k^{'}_i k^{'}_j\zeta(\bm k')\zeta(\bm k-\bm k')\frac{1}{k^2}I^{A}(k,u,v,x),
\end{equation}
with the integral kernel
\begin{equation}
\begin{split}
\label{I_int}
I^{A}(k,u,v,x)&=\int_0^x\mathrm{d}\bar{x}\frac{a(\bar{\tau})}{a(\tau)}k G^A_{k}(\tau,\bar{\tau})f_{\text{PC}}(u,v,\bar x).
\end{split}
\end{equation}

Combining the expressions of scale factor $a$, the Green's function \eqref{GREEN} and $f_{\text{PC}}$ \eqref{fpc1} and \eqref{fpc2}, we obtain
\begin{equation}
I^A(u,v,x)=\frac{\sin x}{x}\mathcal{I}_s(u,v,x)+\frac{\cos x}{x}\mathcal{I}_c(u,v,x),
\end{equation}
where
\begin{equation}
\mathcal{I}_s(u,v,x)=\int^x_0 d \bar{x} f_{\text{PC}}(u,v,\bar{x})\cos \bar{x},
\end{equation}
and
\begin{equation}
\mathcal{I}_c(u,v,x)=-\int^x_0 d \bar{x} f_{\text{PC}}(u,v,\bar{x})\sin \bar{x}.
\end{equation}
We have used the approximation $\omega_A\simeq k$ for simplicity because the deviation of speed of GWs from the speed of light is very small. At late-time, 
\begin{equation}
I^A(u,v,x \rightarrow \infty)^2=\frac{1}{x^2}\left(\sin^2 (x) \mathcal{I}_s(u,v)^2+\cos^2 (x) \mathcal{I}_c(u,v)^2+\sin (2x) \mathcal{I}_s(u,v)\mathcal{I}_c(u,v)\right),
\end{equation}
with
\begin{equation}
\mathcal{I}_s(u,v)=\lim_{x\rightarrow\infty}\mathcal{I}_s(u,v,x),\qquad \mathcal{I}_c(u,v)=\lim_{x\rightarrow\infty}\mathcal{I}_c(u,v,x).
\end{equation}

Then using $\overline{\sin^2 (x)}=\overline{\cos^2 (x)}=1/2$, and $\overline{\sin (2x)}=0$, we obtain the time averaged kernel
\begin{equation}
\overline{I^A(u,v,x \rightarrow \infty)^2}=\frac{1}{x^2}\frac{1}{2}\left(\mathcal{I}_s(u,v)^2+\mathcal{I}_c(u,v)^2\right),
\end{equation}
where the overline represents the time average. For later convenience,  we define a time-independent kernel, $\overline{\tilde{I}^{A}(k,u,v,x)^2}=\overline{I^{A}(k,u,v,x)^2}x^2$ at late-time.

The power spectra of the SIGWs $\mathcal{P}_{h}^{A}$ are defined by 
\begin{equation}
\langle h^A_{\bm{k}} h^C_{\bm{k}'}\rangle =\frac{2\pi^2}{k^3}\delta^3(\bm k+\bm k')\delta^{AC}\mathcal{P}^{A}_{h}(k).
\end{equation}
Combining the definition of $\mathcal{P}_{h}^{A}$ with the solution of SIGWs, we obtain the expression of the power spectra of SIGWs as
\begin{align}
\mathcal{P}^{A}_h(k,x)=4\int_{0}^\infty\mathrm{d}u\int_{|1-u|}^{1+u}\mathrm{d}v
\mathcal{J}(u,v)I^{A}(u,v,x)^2\mathcal{P}_\zeta(uk)\mathcal{P}_\zeta(vk),
\end{align}
where
\begin{equation}
\mathcal{J}(u,v)=\left[\frac{4u^2-(1+u^2-v^2)^2}{4uv}\right]^2,
\end{equation}
and the power spectrum of primordial curvature perturbation is defined by 
\begin{equation}
\langle \zeta_{\bm{k}} \zeta_{\bm{k}'}\rangle =\frac{2\pi^2}{k^3}\delta^3(\bm k+\bm k')\mathcal{P}_{\zeta}(k).
\end{equation}

The fractional energy density of the SIGWs is given by 
\begin{equation}\label{OGW}
\begin{split}
\Omega_{\mathrm{GW}}(k,x)&=\frac{1}{12}\left(\frac{k}{\mathcal{H}}\right)^2\sum\limits_{A=R,L}\overline{\mathcal{P}^A_h(k,x)}=\frac{x^2}{12}\sum\limits_{A=R,L}\overline{\mathcal{P}^A_h(k,x)}\\
&=\frac{1}{3}\int_{0}^\infty\mathrm{d}u\int_{|1-u|}^{1+u}\mathrm{d}v
\mathcal{J}(u,v)\sum\limits_{A=R,L}\overline{\tilde{I}^{A}(k,u,v,x)^2}\mathcal{P}_\zeta(uk)\mathcal{P}_\zeta(vk).
\end{split}
\end{equation}

\acknowledgments

The author Fengge Zhang thanks Junjie Zhao and Yizhou Lu for helpful discussions. This work was supported by National Natural Science Foundation of China under the Grants No. 12305075, the Startup Research Fund of Henan Academy of Sciences under Grants number 241841223, and Joint Fund for Scientific and Technological Research of Henan Province under Grants number 235200810101.

\providecommand{\href}[2]{#2}\begingroup\raggedright\endgroup

% \bibliographystyle{JHEP}
% \bibliography{main}

\end{document}